\documentclass[preprint,12pt]{elsarticle}

\providecommand{\bst}[7]{#1, #7, #2 #3 (#4) #6.} %
\providecommand{\au}[2]{#1 #2} %
\providecommand{\ea}{et al.}

\providecommand{\AP}{Ann. Phys.} %
\providecommand{\ApJ}{ApJ} %
\providecommand{\ApP}{Astropart. Phys.} %
\providecommand{\ARNPS}{Ann. Rev. Nucl. Part. Sci.} %
\providecommand{\CTP}{Commun. Theor. Phys.} %
\providecommand{\GCN}{GCN Circ.} %
\providecommand{\GRG}{Gen. Rel. Grav.} %
\providecommand{\IJMPA}{Int. J. Mod. Phys. A} %
\providecommand{\IJMPD}{Int. J. Mod. Phys. D} %
\providecommand{\JCAP}{J. Cosmol. Astropart. Phys.} %
\providecommand{\LRR}{Living Rev. Rel.} %
\providecommand{\Nat}{Nature} %
\providecommand{\PLB}{Phys. Lett. B} %
\providecommand{\PRD}{Phys. Rev. D} %
\providecommand{\PRL}{Phys. Rev. Lett.} %
\providecommand{\RMP}{Rev. Mod. Phys.} %
\providecommand{\SCI}{Science} %

\journal{Astroparticle Physics}

\begin{document}

\begin{frontmatter}

\title{Lorentz violation from cosmological
objects with very high energy photon emissions}

\author{Lijing Shao}
\author{Zhi Xiao}
\author{Bo-Qiang Ma}\ead{mabq@phy.pku.edu.cn}
\address{School of Physics and State Key Laboratory
of Nuclear Physics and Technology,
\\Peking University, Beijing 100871, China}

\begin{abstract}
Lorentz violation (LV) is predicted by some quantum gravity
theories, where photon dispersion relation is modified, and the
speed of light becomes energy-dependent. Consequently, it results in
a tiny time delay between high energy photons and low energy ones.
Very high energy (VHE) photon emissions from cosmological distance
can amplify these tiny LV effects into observable quantities. Here
we analyze four VHE $\gamma$-ray bursts (GRBs) from Fermi
observations, and briefly review the constraints from three TeV
flares of active galactic nuclei (AGNs) as well. One step further,
we present a first robust analysis of VHE GRBs taking the intrinsic
time lag caused by sources into account, and give an estimate to
quantum gravity energy $\sim 2 \times 10^{17}$~GeV for linear energy
dependence, and $\sim 5 \times 10^9$~GeV for quadratic dependence.
However, the statistics is not sufficient due to the lack of data,
and further observational results are desired to constrain LV
effects better.
\end{abstract}


\end{frontmatter}

\clearpage

\section{Introduction}

The unification of standard model and general relativity is the most
intriguing and desirous goal of modern physics, and it stimulates
the development of many theoretical ideas such as string theory and
loop gravity in the past decades. It is interesting that some of
them predict Lorentz symmetry violation (LV) at low energies or
being realized in a highly nonlinear form. This has arisen in the
space-time foam~\cite{am97,am98,emn99,emn08}, loop
gravity~\cite{gp99,amu00}, torsion in general gravity~\cite{Yan},
vacuum condensate of antisymmetric tensor fields in string
theory~\cite{ks89,ks91,ck97,ck98,arnps09}, and the so-called double
special relativity~\cite{ame02c,ame02,ame02b}.

In most of LV theories, the photon dispersion relation is modified
and the correction is believed to be suppressed by the large Planck
scale, $E_{\rm P} \equiv \sqrt{\hbar c^5 /G} \simeq 1.22 \times
10^{19}$~GeV~\cite{jlm06,km09}. These theories are mainly classified
into two categories. One is effective field theory (EFT), which
provides an excellent framework where the tiny LV effects are
introduced through LV operators: renormalizable ones with dimension
three and/or four, see, e.g., standard model extension
(SME)~\cite{ck97,ck98}, and the further extended non-renormalizable
ones with dimension five and/or six~\cite{arnps09,mp03}. However,
not all quantum gravity theories can be embedded into the EFT
framework, such as the quantum space-time foam
model~\cite{am97,am98,emn99,emn08} and double special
relativity~\cite{ame02c,ame02,ame02b}, and they also introduce
modifications to the canonical dispersion relation. Consequently,
they can result in an energy dependence of the speed of light in the
vacuum, owing to the propagation through an effective gravitational
medium containing space-time quantum fluctuations. Fortunately, this
scenario of lower energy ``relic probe" of Planck scale events can
be tested carefully through laboratory experiments~\cite{ge,am09} or
astronomical
observations~\cite{el00,jlm03,m05,gs08,Xiao:2008yu,Bi:2008yx,as09,emn09}.

Generally, the most model-independent photon dispersion relation
reads in the context of Taylor series as,
\begin{equation}\label{ve}
v(E) = c_0 \left( 1 - \xi \frac{E}{E_{\rm P}} - \zeta
\frac{E^2}{E_{\rm P}^2} \right) \, ,
\end{equation}
where $v(E)$ is the speed of photons with energy $E$, $c_0$ is the
speed of low energy photons, and $\xi$, $\zeta$ are model-dependent
parameters, characterizing the energy where LV occurs. Because of
the suppression of Planck energy, the terms of higher orders are
negligible, and the quadratic term takes effect only when the linear
term vanishes. It could be possible that a more concrete form of
dispersion relation may contain more specific terms of energy
dependence, but we may take Eq.~(\ref{ve}) as a general estimate in
a model-independent manner.

\section{High energy $\gamma$-ray bursts}

Amelino-Camelia~\ea~\cite{am97,am98} first suggested using
cosmological $\gamma$-ray bursts (GRBs) to test LV.  Due to the
large cosmological distance and the fine time structure of GRBs,
tiny LV effects can be amplified into observable quantities. By
taking into account cosmological expansion of the universe, the time
lag induced by LV modified dispersion relation, i.e.,
Eq.~(\ref{ve}), between photons with high energy $E_{\rm h}$, and
those with low energy $E_{\rm l}$, is~\cite{jp08},
\begin{equation}\label{lag}
\Delta t_{\rm LV} = \frac{1+n}{2 H_0} \left( \frac{E_{\rm h}^n -
E_{\rm l}^n}{E_{\rm QG}^n} \right) \int_0^z
\frac{(1+z^\prime)^n {\rm d} z^\prime}{h(z^\prime)} \,,
\end{equation}
where $n=1$ and $n=2$ stand for linear and quadratic energy
dependence, with quantum gravity energy $E_{\rm QG,L} = |\xi|^{-1}
E_{\rm P}$ and $E_{\rm QG,Q} = |\zeta|^{-1/2} E_{\rm P}$,
respectively; $H_0 \simeq 71$~km~s$^{-1}$~Mpc$^{-1}$ is the Hubble
constant; $z$ is the redshift of the source, and $h(z)$ is defined
as
\begin{equation}
h(z) = \sqrt{\Omega_{\rm \Lambda}  + \Omega_{\rm M} (1+z)^3 } \, ,
\end{equation}
where $\Omega_{\rm \Lambda} \simeq 0.73$ is the vacuum energy
density, and $\Omega_{\rm M} \simeq 0.27$ is the matter energy
density in current universe.

Due to the launch of high quality satellites, e.g., Swift and Fermi,
our understanding of GRBs has accomplished evolutionary
improvements. Especially, Fermi Large Area Telescope (LAT), which is
sensitive up to energy of photons $\sim 300$~GeV, has led us to the
very high energy (VHE) domain and inaugurates a new era. LAT
discovered that high energy photons have a tendency to arrive later
relative to low energy ones, which might present potential evidence
for LV~\cite{as09,ab09,ab09b}.

However, the determination of time lag from observational data is
highly nontrivial and affected by many facets, both artificial and
instrumental. As a case study of GRB~090510, Ref.~\cite{ab09b}
discussed several choices, e.g., the time lag between the arrival of
the highest energy photon and the Gamma-ray Burst Monitor (GBM)
trigger, the onset of the main GBM emission, the onset of $>0.1$~GeV
emission, and the onset of $>1$~GeV emission. For simplicity, we
here choose the observed time lag $\Delta t_{\rm obs}$ as the
difference between the arrival of the highest energy photon and the
GBM trigger time. The energy of GBM trigger photons is about
$0.1$~MeV, therefore it is negligible in Eq.~(\ref{lag}), compared
to the highest energy photons, whose energies are significantly
larger than $\sim 1$~GeV.

Four delayed GRBs with known redshifts, observed by the LAT
instrument, are listed in Table~\ref{grb}. Their indicated LV scales
are given as well, relying on the assumption $\Delta t_{\rm obs} =
\Delta t_{\rm LV}$.
As the central engines and emission mechanism of GRBs are not
totally understood yet, and the lags have been detected explicitly
in serval events, we boldly treat the results as possible indicators
of LV effects instead of lower boundaries.

The values derived in the last two columns of Table~\ref{grb} differ
more than tenfold between each other, with average values $E_{\rm
QG,L} \sim (4.9 \pm 8.1) \times 10^{18}$~GeV and $E_{\rm QG,Q} \sim
(1.4 \pm 1.3) \times 10^{10}$~GeV utilizing the least square method.
We attribute the large deviations to the fact that all source
effects are neglected here.

\begin{table}
\begin{center}
\caption{The time lag of the highest energy photons of Fermi LAT
GRBs, relative to the Fermi GBM trigger time. The possible Lorentz
violation energies, $E_{\rm QG,L}$ and $E_{\rm QG,Q}$, are listed for
linear and quadratic energy dependence respectively, without
astrophysical effects taking into account.}\label{grb}
\begin{tabular}{ccccccc}
\hline\hline
GRBs & $z$ & $E$~(GeV) & $\Delta t_{\rm obs}$~(s) & $E_{\rm QG,L}$~(GeV) & $E_{\rm QG,Q}$~(GeV)\\
\hline %
080916C~\cite{ab09} & $4.35$~\cite{gr09} & $13.22$ &
$16.54$ & $1.5 \times 10^{18}$ & $9.7 \times 10^9$ \\
090510~\cite{ab09b} & $0.903$~\cite{ra09} & $31$ & $0.829$ & $1.7 \times 10^{19}$ & $3.4 \times 10^{10}$ \\
090902B~\cite{pbt09} & $1.822$~\cite{cu09} & $33.4$ & $82$ & $3.7 \times 10^{17}$ & $5.9 \times 10^9$ \\
090926A~\cite{utm09} & $2.1062$~\cite{ma09} & $19.6$ & $26$ & $7.8 \times 10^{17}$ & $6.8 \times 10^9$ \\
\hline %
\end{tabular}
\end{center}
\end{table}

The primary uncertainty comes from the unknown effects from source
activities, mainly due to our imperfect knowledge of radiation
mechanism of GRBs. However, we can separate the source effects if we
can achieve a survey of GRBs at different redshifts. The time lag
induced by LV accumulates with propagation distance, as it is a
gravitational medium effect. On the contrary, the intrinsic source
induced time lag is likely to be a distance-independent quantity,
which can be regarded as a constant for a particular class of
sources in the leading order approximation.

Ellis~\ea~\cite{el06,el08} have led a robust analysis of sets of
GRBs from BATSE, HETE, and Swift, utilizing the wavelet
technique~\cite{el03} to search for potential lags. However, due to
the scarcity of the VHE observational data then, no survey of VHE
GRBs with explicit time delay has been treated in a robust way yet.
We here make a first coarse attempt to include available LAT GRBs
and give a global estimate to LV parameters.

On assuming that the intrinsic time lag $\Delta t_{\rm in}$,
originated from astrophysical effects, is independent of redshift
and constant for objects of a particular class, which depends only
on the type of sources, then the observed delay is
\begin{equation}\label{tlag}
\Delta t_{\rm obs} = \Delta t_{\rm LV} + \Delta t_{\rm in} (1+z) \,.
\end{equation}
Inspired by Refs.~\cite{el06,el08}, after a few steps from
Eq.~(\ref{lag}) and Eq.~(\ref{tlag}), we can get a linear formula
with an intercept $\Delta t_{\rm in}$,
and the slope of the line equals to $1/E_{\rm QG,L}$ for the linear
energy dependence and $1/E_{\rm QG,Q}^2$ for the quadratic
dependence,
\begin{equation}
\Delta t_{\rm obs}/(1+z) = K/E_{\rm QG}^n
+ \Delta t_{\rm in} \,,
\end{equation}
where $K$ is defined as
\begin{equation}
K = \frac{1+n}{2 H_0} \frac{E_{\rm h}^n -
E_{\rm l}^n}{1+z} \int_0^z
\frac{(1+z^\prime)^n {\rm d} z^\prime}{h(z^\prime)} \,.
\end{equation}

\begin{figure}
\begin{center}
\includegraphics[width=9cm]{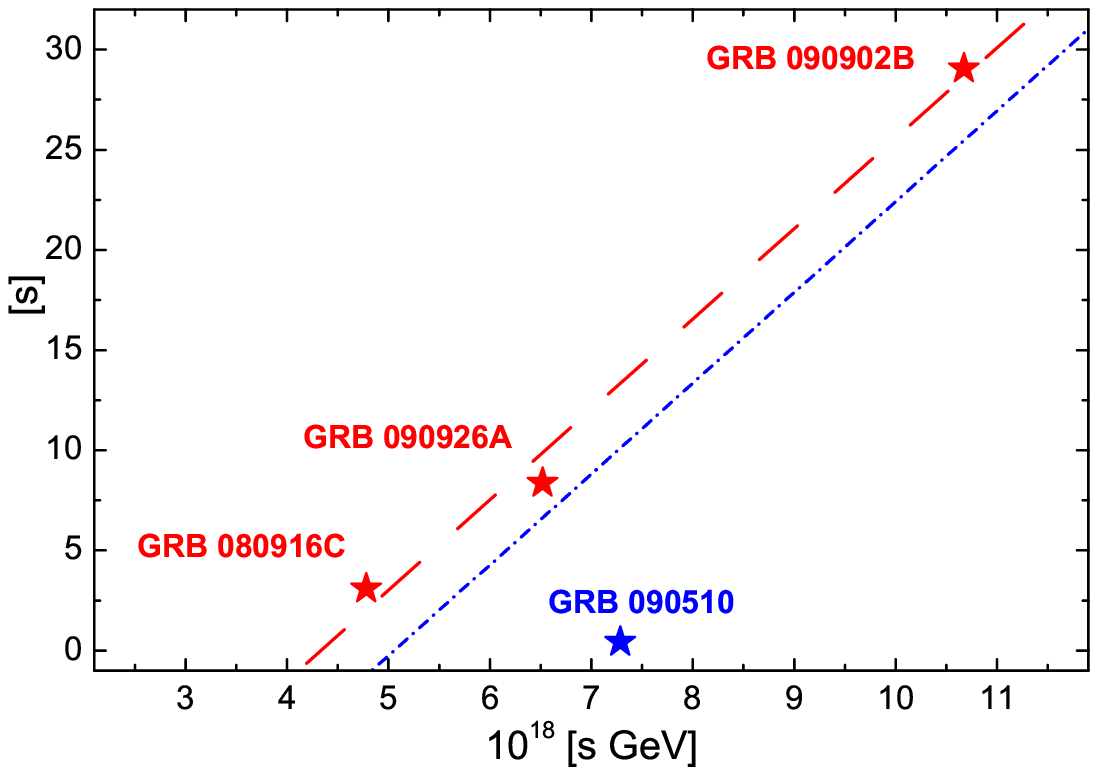}\caption{Linear fits to Fermi LAT GRBs
where linear energy-dependent LV effects and constant intrinsic
source effects are considered. The dash-dotted line stands for a fit
to all four GRBs, while the dashed line only fits to three long GRBs
(without the short GRB~090510). The intercept and slope of fitted
line equal to $\Delta t_{\rm in}$ and $1/E_{\rm QG,L}$,
respectively.\label{lgrb}}
\end{center}
\end{figure}

\begin{figure}
\begin{center}
\includegraphics[width=9cm]{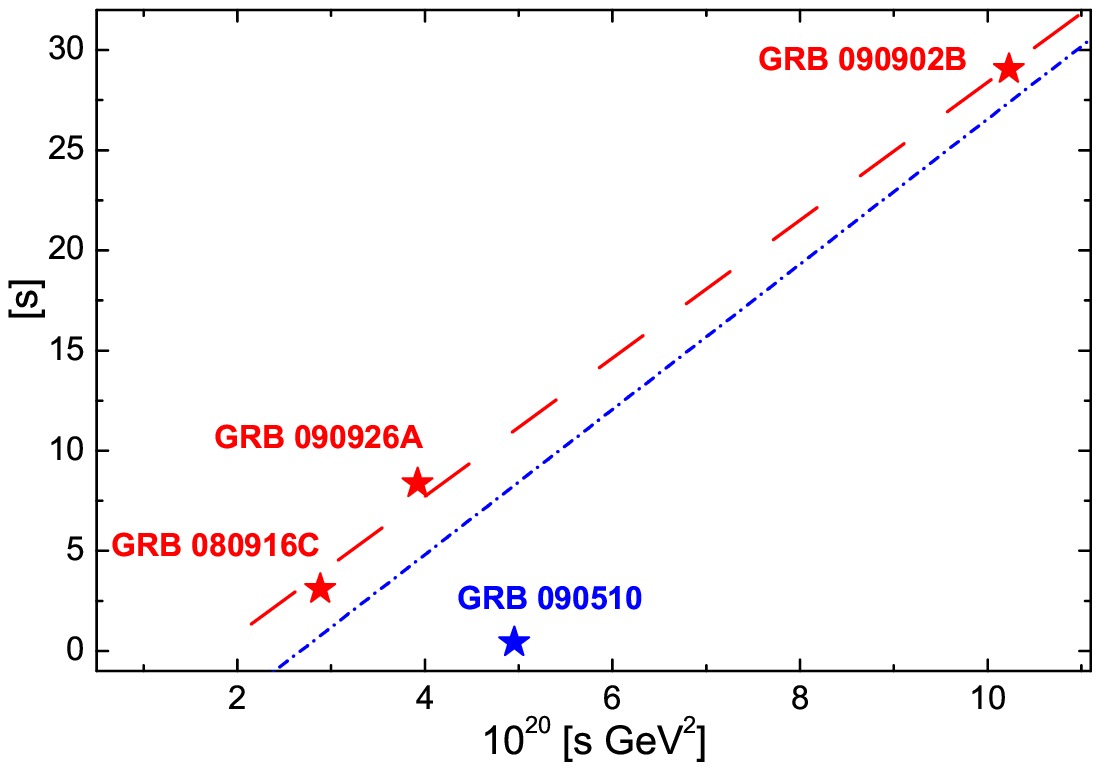}\caption{Linear fits to Fermi LAT GRBs
where quadratic energy-dependent LV effects and constant intrinsic
source effects are considered. The dash-dotted line stands for a fit
to all four GRBs, while the dashed line only fits to three long GRBs
(without the short GRB~090510). The intercept and slope of fitted
line equal to $\Delta t_{\rm in}$ and $1/E_{\rm QG,Q}^2$,
respectively.\label{qgrb}}
\end{center}
\end{figure}

The plots of linear fits for two energy-dependent scenarios are
illustrated in Fig.~\ref{lgrb} and Fig.~\ref{qgrb} individually. The
dash-dotted lines represent linear fit to all four GRBs from
Table~\ref{grb}, while the dashed lines only fit to three long GRBs
(GRB~080916C, GRB~090902B, and GRB~091003A; duration $T_{90}>2$~s).
With the consideration that short GRB~090510 (duration $T_{90}<2$~s)
would have different intrinsic time delay from the long ones due to
their distinct progenitor mechanism~\cite{nb06,p05}, we expect that
these two classes of GRBs should have different intercepts if more
data of short GRBs are available. Actually,
current prevailing paradigm regards that long GRBs come from the
collapses of massive rapidly rotating stars, while  short GRBs are
believed to be originating from the coalescence of two neutron stars
or a neutron star and a black hole~\cite{p05}. Thus even with
absence of more data, the apparent deviation of GRB~090510 from the
red dashed line is expected.

Due to the lack of statistics, we do not give the errors of data as
it is still too early to constrain LV parameters accurately without
high enough statistics. Therefore, one should caution that the data
are rather rough and strongly depend on artificial choices as
mentioned. Nevertheless, from the figures, we can see that when only
three long GRBs are included, the linear and quadratic fits are both
very likely and give quantum gravity energy as $E_{\rm QG,L} =
(2.2\pm0.2) \times 10^{17}$~GeV and $E_{\rm QG,Q} =
(5.4\pm0.2) \times 10^9$~GeV. After including the short
GRB~090510, the fits are still reasonable and give $E_{\rm QG,L} =
(2.2\pm0.9) \times 10^{17}$~GeV and $E_{\rm QG,Q} =
(5.3\pm0.8) \times 10^9$~GeV. We can see that the mean values
almost stay the same for two choices, whereas the fitted errors are
somehow larger in the later fit due to inclusion of short
GRB~090510. Our procedure avoids the inconsistently large deviations
derived from different sources without taking astrophysical effects
into account, as shown in the last two columns of Table~\ref{grb}.

Actually, the method to determine quantum gravity scale, when only
one source is included without considering the intrinsic effects, is
equivalent to only using the slope between a specific data point and
the point of origin in our figures. Hence the lower right position
of GRB~090510 in figures gives its slope extremely small compared to
the other three sources, and hence the derived quantum gravity energy,
inverse of the slope, is rather large~\cite{ab09b}. From this
viewpoint, the conclusion that dispersion relation with linear
energy dependence is ruled out because of the early onset of high
energy photons of GRB~090510~\cite{ab09b} seems too early to be
solidified~\cite{xm09}.

\section{Active galactic nuclei}

Active galactic nuclei (AGNs) locate nearer and their time
structures are not so variable as those of GRBs. However, their VHE
emissions are in the TeV range, which is significantly higher than
the highest energy emissions observed in GRBs. Thus they represent
another kind of VHE astronomical laboratories to probe possible
evidence of LV.

Here we briefly review the three AGNs which were published
previously to test LV effects. We recalculate the potential LV
scales utilizing Eq.~(\ref{lag}) and discuss the hints and
limitations from AGNs as well.

{\bf Markarian~421}. It was reported 10 years ago, that no time lag
larger than $280$~s was found between energy bands $< 1$~TeV and
$>2$~TeV during a TeV flare of Markarian~421~\cite{bi99}. The AGN is
known to locate at $z = 0.031$, thus it sets a lower boundary to LV
with quantum gravity energy scales $E_{\rm QG,L}> 4.9 \times
10^{16}$~GeV and $E_{\rm QG,Q}> 1.5 \times 10^{10}$~GeV.

{\bf Markarian~501}. MAGIC Collaboration found a time lag about
$4$~min for photons in the 1.2--10~TeV energy band relative to those
in the range 0.25--0.6~TeV during a VHE flare of
Markarian~501~\cite{al08}, whose redshift equals to 0.034. The mean
difference of the two bands is reported $\approx 2$~TeV, thus we can
do an estimation and obtain $E_{\rm QG,L} \sim 1.2 \times
10^{17}$~GeV, regardless of the intrinsic delay. This value is very
close to our results from the global fits of GRBs.

{\bf PKS~2155-304}. HESS Collaboration published an interesting VHE
flare of the BL Lacertae object PKS~2155-304~\cite{ah08}, which
locates at $z=0.116$~\cite{fpt93}, more distant than Markarian~421
and Markarian~501. They made use of the modified cross correlation
function and got a time lag $\sim 20$~s between lightcurves of two
different energy bands. However, the delay is not sufficiently
significant. They reported that the mean difference of two energy
bands in this case is $1.0$~TeV, while the mean quadratic difference
is $2.0$~TeV$^2$, hence we get the potential LV scales to be $E_{\rm
QG,L} \sim 2.6 \times 10^{18}$~GeV and $E_{\rm QG,Q} \sim 9.1 \times
10^{10}$~GeV accordingly. They are both about one magnitude larger
than our robust values.

It is worthy to mention that the estimations above are based on the
assumption $\Delta t_{\rm in} = 0$, and because of ignorance of the
source mechanism, they remain controversial~\cite{as09,emn09}. And
global fits for AGNs of different types are expected to present
something different, like what happens in the GRBs case. However,
current AGNs data are inadequate to carry out a robust analysis.
Moreover, TeV flares from AGNs seem relatively rare and
unpredictable, and they are produced only occasionally by AGNs with
restricted redshifts contrary to large and diverse redshifts of
GRBs~\cite{emn09}. Furthermore, the sources of AGNs are very
different, and we expect they have distinct astrophysical lags and
intrinsic fluctuations. Therefore, it is not likely to present a
robust analysis of AGNs for LV effects, at least in the recent
future. However, they provide a complementary probe to LV effects
due to the different observational method and distinct origins.

\section{Discussions and summary}

Nowadays, more very high energy (VHE) data become available and they
provide a rich ground to look into cosmologically accumulated
effects, rooted from quantum gravity scale. In this paper, we use
cosmological objects such as $\gamma$-ray bursts (GRBs) and active
galactic nuclei (AGNs) to estimate Lorentz violation (LV) effects
from the time delay between photons with different energies. The
quantum gravity scales from GRBs and AGNs are surprisingly
compatible in some sense.

We give a crude attempt here to survey a set of VHE celestial events
for LV hints and disentangle astrophysical origins from LV effects
as well. Though the data are very limited in our study, even not
sufficient to establish a good statistics, we get some potential
clues of LV from these samples. Further, we notice that more
observations are emerging and more experiment data are accumulating,
e.g., Fermi LAT has detected GRB~090323, GRB~090328 and GRB~091003
with VHE emissions and their redshifts are attained from other
observations as well, thus once these results are published, there
will definitely be an opportunity to give a more stringent analysis.
We suggest combining VHE GRBs at different redshifts to separate
source effects, if data points approximately lay on two lines
according to two different GRB types, it will be a supportive
evidence to our approach.

Although the results in our analysis are very preliminary, they give
hints for possible LV energy scales, $\sim 2 \times 10^{17}$~GeV for
linear energy dependence and $\sim 5 \times 10^9$~GeV for quadratic
dependence. However, the fits in Fig.~\ref{lgrb} and Fig.~\ref{qgrb}
induce a negative intercept, $\Delta t_{\rm in} \sim -20$~s for the
linear dependence and $-6$ to $-10$~s for the quadratic scenario,
which means that at sources, high energy photons emit earlier than
low energy ones. This conflicts with common expectations based on
assumptions that low energy photons are produced by electrons while
high energy ones are generated by protons, and because of lighter
masses, electrons are accelerated earlier and hence low energy
photons come out first~\cite{emn09}. However, because that the last
word about the emission mechanism of GRBs is not addressed yet, and
the negative intercepts are not very significant in our fits, they
may not trouble much.

Worthy to mention that, there are several stringent restrictions
coming from various tests on LV effects in the content of given
theories~\cite{arnps09,jlm06,m05,as09,emn09}. Synchrotron radiation
measurements on electrons from the Crab Nebula~\cite{jlm03,m07}
place very strong constraints on the electron sector, making the
linear dependence for electrons almost impossible. Most effective
field theories (EFTs) predict birefringence for the photon
sector~\cite{gp99,mp03}, and astrophysical observation leads to very
tight restrictions on the linear suppression~\cite{gk01}. And the
GZK cutoff originated from the ultra-high-energy cosmic rays
(UHECRs) interacting with cosmic microwave background (CMB) photons
also produces severe constraints for linear as well as quadratic
energy dependence~\cite{gs08,Xiao:2008yu,Bi:2008yx}. However, some
scenarios can avoid the above restrictions~\cite{emn09}, e.g.,
photons have different LV parameters as electrons and hadrons, thus
our analysis above serves as a tentative estimate for some still
surviving theories.

Finally, we stress that it would be premature to draw a rigorous
conclusion on LV at the moment, and more theoretical considerations
and practical data analysis are needed for a more stringent
constraint and clarification on these issues.

\section*{Acknowledgements}

This work is supported by National Natural Science Foundation of
China (Nos. 10721063, 10975003), and National Fund for Fostering
Talents of Basic Science (Nos. J0630311, J0730316). It is also
supported by Hui-Chun Chin and Tsung-Dao Lee Chinese Undergraduate
Research Endowment (Chun-Tsung Endowment) at Peking University.

\bibliography{99}

\begin{thebibliography}{99}

\bibitem{am97}
\bst{\au{G.}{Amelino-Camelia}~\ea}{\IJMPA}{12}{1997}{607}{607-623}{Distance
measurement and wave dispersion in a Liouville-string approach to
quantum gravity}

\bibitem{am98}
\bst{\au{G.}{Amelino-Camelia}~\ea}{\Nat}{393}{1998}{763}{763-765}{Tests
of quantum gravity from observations of $\gamma$-ray bursts}

\bibitem{emn99}
\bst{\au{J.}{Ellis}, \au{N.E.}{Mavromatos},
\au{D.V.}{Nanopoulos}}{\GRG}{31}{1999}{1257}{1257-1262}{Search for
quantum gravity}

\bibitem{emn08}
\bst{\au{J.}{Ellis}, \au{N.E.}{Mavromatos},
\au{D.V.}{Nanopoulos}}{\PLB}{665}{2008}{412}{412-417}{Derivation of
a vacuum refractive index in a stringy space-time foam model}

\bibitem{gp99}
\bst{\au{R.}{Gambini},
\au{J.}{Pullin}}{\PRD}{59}{1999}{124021}{124021}{Nonstandard optics
from quantum space-time}

\bibitem{amu00}
\bst{\au{J.}{Alfaro}, \au{H.A.}{Morales-T\'ecotl},
\au{L.F.}{Urrutia}}{\PRL}{84}{2000}{2318}{2318-2321}{Quantum gravity
corrections to neutrino propagation}

\bibitem{Yan}
\bst{\au{M.L.}{Yan}}{\CTP}{2}{1983}{1281}{1281-1296}{The
renormalizability of general gravity theory with torsion and the
spontaneous breaking of Lorentz group}

\bibitem{ks89}
\bst{\au{V.A.}{Kosteleck\'y},
\au{S.}{Samuel}}{\PRL}{63}{1989}{224}{224-227}{Phenomenological
gravitational constraints on strings and higher-dimensional
theories}

\bibitem{ks91}
\bst{\au{V.A.}{Kosteleck\'y},
\au{S.}{Samuel}}{\PRL}{66}{1991}{1811}{1811-1814}{Photon and
graviton masses in string theories}

\bibitem{ck97}
\bst{\au{D.}{Colladay},
\au{V.A.}{Kosteleck\'y}}{\PRD}{55}{1997}{6760}{6760-6774}{CPT
voilation and the standard model}

\bibitem{ck98}
\bst{\au{D.}{Colladay},
\au{V.A.}{Kosteleck\'y}}{\PRD}{58}{1998}{116002}{116002}{Lorentz-violating
extension of the standard model}

\bibitem{arnps09}
\bst{\au{S.}{Liberati},
\au{L.}{Maccione}}{\ARNPS}{59}{2009}{245}{245-267}{Lorentz
violation: motivation and new constraints}

\bibitem{ame02c}
\bst{\au{G.}{Amelino-Camelia}}{\IJMPD}{11}{2002}{35}{35-60}{Relativity
in space-times with short-distance structure governed by an
observer-independent (Planckian) length scale}

\bibitem{ame02}
\bst{\au{G.}{Amelino-Camelia}}{\Nat}{418}{2002}{34}{34-35}{Relativity:
special treatment}

\bibitem{ame02b}
\bst{\au{G.}{Amelino-Camelia}}{\IJMPD}{11}{2002}{1643}{1643-1669}{Doubly-special
relativity: first results and key open problems}

\bibitem{jlm06}
\bst{\au{T.}{Jacobson}, \au{S.}{Liberati},
\au{D.}{Mattingly}}{\AP}{321}{2006}{150}{150-196}{Lorentz violation
at high energy: concepts, phenomena, and astrophysical constraints}

\bibitem{km09}
\bst{\au{V.A.}{Kosteleck\'y},
\au{M.}{Mewes}}{\PRD}{80}{2009}{015020}{015020}{Electrodynamics with
Lorentz-violating operators of arbitrary dimension}

\bibitem{mp03}
\bst{\au{R.C.}{Myers},
\au{M.}{Pospelov}}{\PRL}{90}{2003}{211601}{211601}{Ultraviolet
modifications of dispersion relations in effective field theory}

\bibitem{ge}
\bst{\au{H.}{M${\rm
\ddot{u}}$ller}~\ea}{\PRL}{99}{2007}{050401}{050401}{Tests of
relativity by complementary rotating Michelson-Morley experiments}

\bibitem{am09}
\bst{\au{G.}{Amelino-Camelia}~\ea}{\PRL}{103}{2009}{171302}{171302}{Constraining
the energy-momentum dispersion relation with Planck-scale
sensitivity using cold atoms}

\bibitem{el00}
\bst{\au{J.}{Ellis}~\ea}{ApJ}{535}{2000}{139}{139-151}{A search in
gamma-ray burst data for nonconstancy of the velocity of light}

\bibitem{jlm03}
\bst{\au{T.}{Jacobson}, \au{S.}{Liberati},
\au{D.}{Mattingly}}{\Nat}{424}{2003}{1019}{1019-1021}{A strong
astrophysical constraint on the violation of special relativity by
quantum gravity}

\bibitem{m05}
\bst{\au{D.}{Mattingly}}{\LRR}{8}{2005}{5}{5-84}{Modern tests of
Lorentz invariance}

\bibitem{gs08}
\bst{\au{M.}{Galaverni},
\au{G.}{Sigl}}{\PRL}{100}{2008}{021102}{021102}{Lorentz violation
for photons and ultrahigh-energy cosmic rays}


\bibitem{Xiao:2008yu}
  Z.~Xiao, B.-Q.~Ma,
  Lorentz violation dispersion relation and its application,
  Int.\ J.\ Mod.\ Phys.\  A {\bf 24} (2009) 1359-1381.




\bibitem{Bi:2008yx}
  X.-J.~Bi, Z.~Cao, Y.~Li and Q.~Yuan,
  Testing Lorentz invariance with the ultra-high energy cosmic ray
  spectrum,
  Phys.\ Rev.\  D {\bf 79} (2009) 083015.


\bibitem{as09}
\bst{\au{G.}{Amelino-Camelia},
\au{L.}{Smolin}}{\PRD}{80}{2009}{084017}{084017}{Prospects for
constraining quantum gravity dispersion with near term observations}

\bibitem{emn09}
\bst{\au{J.}{Ellis}, \au{N.E.}{Mavromatos},
\au{D.V.}{Nanopoulos}}{\PLB}{674}{2009}{83}{83-86}{Probing a
possible vacuum refractive index with $\gamma$-ray telescopes}

\bibitem{jp08}
\bst{\au{U.}{Jacob},
\au{T.}{Piran}}{\JCAP}{01}{2008}{031}{031}{Lorentz-violation-induced
arrival delays of cosmological particles}

\bibitem{ab09}
\bst{\au{A.A.}{Abdo}~\ea}{\SCI}{323}{2009}{1688}{1688-1693}{Fermi
observations of high-energy gamma-ray emission from GRB 080916C}

\bibitem{ab09b}
\bst{\au{A.A.}{Abdo}~\ea}{\Nat}{462}{2009}{331}{331-334}{A limit on
the variation of the speed of light arising from quantum gravity
effects}


\bibitem{gr09}
\bst{\au{J.}{Greiner}~\ea}{A\&A}{498}{2009}{89}{89-94}{The redshift
and afterglow of the extremely energetic gamma-ray burst GRB
080916C}

\bibitem{ra09} \au{A.}{Rau}~\ea, GRB090510: VLT/FORS2 spectroscopic redshift, \GCN~9353 (2009).

\bibitem{pbt09}
\au{F. de}{Palma}, \au{J.}{Bregeon}, \au{H.}{Tajima}, GRB 090902B:
Fermi LAT detection, \GCN~9867 (2009).

\bibitem{cu09}
\au{A.}{Cucchiara}~\ea, GRB 090902B: Gemini-N absorption redshift,
\GCN~9873 (2009).

\bibitem{utm09}
\au{T.}{Uehara}, \au{H.}{Takahashi}, \au{J.}{McEnery}, GRB 090926:
Fermi LAT detection, \GCN~9934 (2009).

\bibitem{ma09}
\au{D.}{Malesani}~\ea, GRB 090926A: VLT/X-shooter redshift,
\GCN~9942 (2009).

\bibitem{el06}
\bst{\au{J.}{Ellis}~\ea}{\ApP}{25}{2006}{402}{402-411}{Robust limits
on Lorentz violation from gamma-ray bursts}

\bibitem{el08}
\bst{\au{J.}{Ellis}~\ea}{\ApP}{29}{2008}{158}{158-159}{Corrigendum
to ``Robust limits on Lorentz violation from gamma-ray bursts''
[Astropart. Phys. 25 (2006) 402]}

\bibitem{el03}
\bst{\au{J.}{Ellis}~\ea}{A\&A}{402}{2003}{409}{409-424}{Quantum-gravity
analysis of gamma-ray bursts using wavelets}

\bibitem{nb06}
\bst{\au{J.P.}{Norris},
\au{J.T.}{Bonnell}}{ApJ}{643}{2006}{266}{266-275}{Short gamma-Ray
bursts with extended emission}

\bibitem{p05}
\bst{\au{T.}{Piran}}{\RMP}{76}{2005}{1143}{1143-1210}{The physics of
gamma-ray bursts}

\bibitem{xm09}
\au{Z.}{Xiao}, \au{B.-Q.}{Ma}, Constraints on Lorentz invariance
violation from gamma-ray burst GRB090510, 
  Phys.\ Rev.\  D {\bf 80} (2009) 116005.




\bibitem{bi99}
\bst{\au{S.D.}{Biller}~\ea}{\PRL}{83}{1999}{2108}{2108-2111}{Limits
to quantum gravity effects on energy dependence of the speed of
light from observations of TeV flares in active galaxies}

\bibitem{al08}
\bst{\au{J.}{Albert}~\ea}{\PLB}{668}{2008}{253}{253-257}{Probing
quantum gravity using photons from a flare of the active galactic
nucleus Markarian 501 observed by the MAGIC telescope}

\bibitem{ah08}
\bst{\au{F.}{Aharonian}~\ea}{\PRL}{101}{2008}{170402}{170402}{Limits
on an energy dependence of the speed of light from a flare of the
active galaxy PKS 2155-304}

\bibitem{fpt93}
\bst{\au{R.}{Falomo}, \au{J.E.}{Pesce},
\au{A.}{Treves}}{\ApJ}{411}{1993}{L63}{L63-L66}{The environment of
the BL Lacertae object PKS 2155-304}

\bibitem{m07}
\bst{\au{L.}{Maccione}~\ea}{\JCAP}{0710}{2007}{013}{013}{New
constraints on Planck-scale Lorentz violation in QED from the Crab
Nebula}

\bibitem{gk01}
\bst{\au{R.J.}{Gleiser},
\au{C.N.}{Kozameh}}{\PRD}{64}{2001}{083007}{083007}{Astrophysical
limits on quantum gravity motivated birefringence}






\end{thebibliography}

\end{document}